\documentclass[11pt]{article}
\usepackage[makeroom]{cancel}
\usepackage{cite}
\usepackage{graphicx}
\usepackage{multirow}
\usepackage[utf8]{inputenc}
\usepackage[english]{babel}
\usepackage{graphicx}
\usepackage{amsmath}
\usepackage{multicol,lipsum,graphicx,float}
\usepackage{color}
\usepackage{caption}
\usepackage{cite}
\usepackage{blindtext}
\usepackage{hyperref}
\usepackage[bottom]{footmisc}
\usepackage{subcaption}
\makeatletter

\makeatletter
\newcommand*{\rom}[1]{\expandafter\@slowromancap\romannumeral #1@}
\makeatother

\begin{document}
\begin{center}
{\Large\bf Higgs Vacuum Stability and Modified Chaotic Inflation}
\\
\vskip .5cm
{
Abhijit Kumar Saha$^{a,}$\footnote{abhijit.saha@iitg.ernet.in},
Arunansu Sil$^{a,}$\footnote{asil@iitg.ernet.in}}\\[3mm]
{\it{
$^a$ Department of Physics, Indian Institute of Technology Guwahati, 781039 Assam, India}
}
\end{center}

\vskip .5cm

\begin{abstract}
The issue of electroweak vacuum stability is studied in presence of a scalar field which participates in modifying the minimal chaotic inflation model. It is shown that the threshold effect on the Higgs quartic coupling originating from the Higgs-inflaton sector interaction can essentially make the electroweak vacuum stable upto the Planck scale. On the other hand we observe that the new physics parameters in this combined framework are enough to provide deviation from the minimal chaotic inflation predictions so as to keep it consistent with recent observation by Planck 2015.
\noindent \hspace{4cm}
\end{abstract}

The discovery of the Higgs boson\cite{Agashe:2014kda,Aad:2012tfa,
Chatrchyan:2012xdj,Giardino:2013bma} undoubtedly establishes the credibility of the Standard Model (SM) 
as a successful theory of fundamental interactions in nature. While the Higgs boson was the only particle in 
the SM that remained to be discovered until recently, its finding does not necessarily indicate the end for the 
hunt of particle physics. On the contrary, the discovery opens up several questions regarding the Higgs sector 
(or Higgs potential) of the SM. In particular the study of the Higgs potential turns out to be quite intriguing 
in view of the fact that the Higgs quartic coupling $\lambda_h$ becomes negative at high energies 
(the SM instability scale $\Lambda_{I}^{\textrm{SM}}\sim 10^{10}$ GeV) indicating a possible instability of the electroweak (EW) vacuum, as beyond 
$\Lambda_I^{\textrm{SM}}$ the Higgs potential becomes unbounded from below or it might have another 
minimum (a true minimum) at a very large field value\cite{Coleman:1973jx,Sher:1988mj}. This poses (the latter possibility) a 
cosmological problem: why the early universe should favor the EW minimum in presence of a deeper one (the true 
minimum) at a large field value? 

The vacuum stability problem of the Higgs potential is intensely tied up with the precise value of the top quark 
mass $m_t$\cite{Branchina:2014usa,Degrassi:2014hoa}. For a certain range of SM Higgs mass ($m_h$) and $m_t$, the problem 
is bit alleviated provided the lifetime of 
the EW vacuum exceeds the age of the Universe or in other words the EW vacuum becomes metastable. The 
current avilable data suggests the EW vacuum as a metastable one\cite{Tang:2013bz,Buttazzo:2013uya,Isidori:2001bm}. 
However the resolution exclusively depends on the precise 
measurement of $m_t$ that may push it to the instability region. One of the possible 
solutions is to introduce new physics in-between EW scale and $\Lambda_I^{\textrm{SM}}$. In view of SM's incompetence 
to resolve some of the issues like dark matter, matter-antimatter asymmetry, neutrino mass and mixings, 
inflation etc, the introduction of new physics is of course a welcome feature. 

In this work, we want to study how the vacuum stability problem gets affected in presence of  inflation sector in the 
early universe. Successful inflation in early universe  is required for structure formation. The inflationary predictions 
should however be consistent with Planck 2015\cite{Ade:2015xua} data. 
During inflation the Higgs field is expected to receive a typical fluctuation with amplitude of 
$\mathcal{O}(H_{\textrm{Inf}})$ where $H_{\textrm{Inf}}$ is the Hubble scale during inflation. So there is a possibility that 
it can be driven from the EW vacuum toward the unstable part of the 
Higgs potential provided $H_{\textrm{Inf}} > \Lambda_I^{\textrm{SM}}$ which is satisfied by most of the large field inflation models\cite{Kobakhidze:2013tn}. 
Now if the effective 
mass of the Higgs boson can be made sufficiently large during inflation ($m_h^{\textrm{eff}}>H_{\textrm{Inf}}$), the Higgs field will 
naturally evolve to origin and 
the problem can be evaded. This large effective mass term can be generated through Higgs-inflaton interaction 
as suggested in \cite{Gross:2015bea} and the dangerous effect on Higgs vacuum stability during inflation 
could be avoided. So an involvement of new physics (a scalar field involved in inflation) is required which can ensure 
Higgs field to remain at origin during inflationary period of the universe. Once the inflation is over, this field can then
fall in the EW minimum as this minimum is close to the origin.

Concerning the model of inflation, the chaotic inflation\cite{Linde:2007fr} with quadratic potential is perhaps the 
simplest scenario. However it is in the verge of being excluded by the Planck 2015 observations\cite{Ade:2015xua}. Several ways 
are there to save this model \cite{Harigaya:2015pea,Evans:2015mta,Saha:2015afa,Kallosh:2010ug,Li:2013nfa,Harigaya:2012pg, Harigaya:2014fca,Nakayama:2013jka}. Idea behind all these is however common. If we can flatten the 
choatic potential dynamically, it can predict correct value of spectral index ($n_s$) and scalar to tensor ratio ($r$). 
One particular approach\cite{Harigaya:2015pea} seems interesting in this context where 
involvement of a second SM singlet scalar field (apart from the one responsible for chaotic inflation) is assumed. The effect 
of this additional scalar is to modify the quadratic potential $V_{\phi}$ to some extent. 

Here we investigate the possibility of using this extra scalar field of the inflation system to take part in resolving the 
Higgs vacuum stability problem. It is shown in \cite{EliasMiro:2012ay,Anchordoqui:2012fq} that involvement of a scalar field
can indeed modify the stability condition of electroweak vacuum in SM provided this singlet field acquires a large vacuum expectation value. 
The threshold effect provides a tree level shift  in $\lambda_h$ at a scale below which this heavy scalar would be 
integrated out. This turns out to be very effective in keeping the Higgs quartic coupling positive upto the scale $M_{P}$.

 Previously, connecting the inflaton and the Higgs sector to solve the vacuum stability problem has been extensively 
 studied in  \cite{Bhattacharya:2014gva}. They have considered hilltop and quartic inflations where 
inflaton ($\phi$) itself plays the role of this singlet as at the end of inflation $\phi$ field gets a large vev. Below its mass scale 
 the inflaton can be integrated out and the higgs quartic coupling gets a shift. The energy scale 
where this threshold effect occurs is therefore fixed by the inflaton mass $m$. Our approach however differs 
from \cite{Bhattacharya:2014gva}.  We employ the chaotic inflation with potential 
$V_{\phi} = m^2 \phi^2/2 $ where $\langle\phi\rangle=0$ at the end of inflation. Following \cite{Harigaya:2015pea} 
another scalar field $\chi$ is introduced where its coupling with $\phi$  provides the required modification so that 
the inflationary predictions fall within the allowed $n_s-r$ region by Planck 2015\cite{Ade:2015xua}. In addition we suggest 
a modification 
where the $\chi$ field has a large vev ($v_{\chi}$) at its true minimum. It is found that the size of the vev is essentially 
unconstrained from inflation data. On the other hand this field can play important role in studying the Higgs 
vacuum stability issue.
Apart from inflation, moduli fields\cite{Ema:2016ehh} and the scalar singlet(s) involved in dark matter \cite{Gonderinger:2009jp,Gonderinger:2012rd,Khan:2014kba,Khoze:2014xha,Mambrini:2016dca,Baek:2012uj,Chao:2012mx,Chen:2012faa}, 
neutrinos \cite{Datta:2013mta,Chakrabortty:2013zja,Baek:2013fsa,Coriano:2014mpa,Mohapatra:2014qva,Haba:2015rha,Bonilla:2015kna,
Haba:2016zbu,Ng:2015eia,Das:2015nwk
,Rose:2015fua,Kobakhidze:2013pya,Khan:2012zw}
 can have effect on the Higgs vacuum stability in many different ways. 

We start by summarizing the features of the chaotic inflation with quadratic potential, 
\begin{align}
V_{\phi}=\frac{1}{2}m^2\phi^2,
\end{align}
where the inflaton $\phi$ is a real scalar field (SM singlet) and $m$ is the mass parameter. With this potential, 
the slow roll happens at a super-Planckian value of the inflaton field $\phi$. The magnitude of $m$ 
is found to be $m\simeq 1.4\times 10^{13}$ GeV in order to be consistent with the observed amount of 
curvature perturbation. This potential yields the magnitude of spectral index $(n_s)$ 
and tensor to scalar ratio ($r$) as $n_s=1-\frac{2}{N_e}\simeq0.967$ and $r=\frac{8}{N_e}\simeq 0.133$ 
where the number of $e$-folds $(N_e)$ is considered to be 60. However from observational point of view, 
Planck 2015\cite{Ade:2015xua} provides an upper-bound on $r\leq 0.11$ which seems to disfavor this minimal model for its 
prediction of large $r$. So a modification is highly 
appreciated to revive the model. 

Following the recent proposal in \cite{Harigaya:2015pea},  we consider a variant of the above potential 
\begin{align}\label{infpot}
V_{\rm{I}}=\frac{1}{2} m^2\phi^2 - \frac{c_1}{2} \phi^2 (\chi^2-v_{\chi}^2)+\frac{\lambda_\chi}{4}(\chi^2-v_{\chi}^2)^2,
\end{align}
where $m, c_1, \lambda_{\chi}$ and $v_\chi$ are real and positive parameters. $\chi$ is another SM singlet scalar field 
which helps flattening the quadratic part of the potential involving $\phi$. We assume a $Z_2$ symmetry under 
which $\phi$ and $\chi$ fields are odd and hence they appear quadratically in the potential. The vev of the $\chi$ 
field in its global minimum is $v_{\chi}$.
 As it will turn out $v_\chi$ does not have almost any impact on the inflationary 
 predictions\footnote{The sole purpose of introducing a $\chi$ vev is to contribute in the Higgs quartic coupling $\lambda_h$ through threshold effect.}.  
Note that we have not considered 
the higher order terms involving $\phi$, $e.g. ~\phi^4$.  The effect of those terms would be destructive in terms 
of the flatness of the potential. Therefore coefficients associated with those higher order terms in $\phi$ are assumed 
to be negligibly small\footnote{Their absence or smallness can be argued in terms of shift symmetry of the $\phi$ field 
\cite{Mukaida:2014kpa} where $m$ serves as the shift symmetry breaking parameter.}. 

Similar to the original chaotic inflation model with $V_\phi$, here also during inflation the $\phi$ field takes super-Planckian value. 
The potential is such that the $\chi$ field receives a negative mass-squared term which depends on the field value of $\phi$. Therefore $\chi$ 
acquires  a large field value due to its coupling (through $c_1$ term)  with the $\phi$ field
\begin{align}
\langle \chi \rangle=\Big(v_{\chi}^2+\frac{c_1}{\lambda_{\chi}} \phi^2\Big)^{1/2}\simeq \sqrt{\frac{c_1}{\lambda_{\chi}}}\phi=\chi_\textrm{I},
\label{chivev}
\end{align} 
$\chi_\textrm{I}$ is however considered to be sub-Planckian which implies $\frac{c_1}{\lambda_\chi}\ll 1$. Once
$\phi$ rolls down to a smaller value, $\langle\chi\rangle$ decreases. When $\phi$ finally settles at origin, $\langle\chi\rangle$ is shifted to its minimum 
$v_{\chi}$. The magnitude of $v_{\chi}$ is assumed to be  below $m$ $(v_\chi<m)$ so that it does not disturb  
 predictions\cite{Harigaya:2015pea} of the modified chaotic inflation model.

As long as the field $\chi$ is stucked at $\chi_\textrm{I}$ during inflation, its mass is found to satisfy
\begin{equation}\label{chiv}
m_{\chi}^2(\phi)=\frac{\partial^2 V}{\partial \chi^2}\Big|_{\langle\chi\rangle}=2 c_1 \phi^2+2 \lambda_{\chi}v_{\chi}^2.
\end{equation} 
It is to be noted that due to the super-Planckian field value of $\phi$ at the beginning and during inflation, 
$m_\chi^2(\phi)$ turns out to be (with suitable $c_1$ and $\lambda_{\chi}$ as we will see) greater than 
$H_{\rm{Inf}}^2 \simeq \frac{m^2 \phi^2} {6 M_P^2}$ where $M_P=2.4\times 10^{18}$ GeV is the reduced Planck scale. Hence $\chi$ is expected to be stabilized at $\chi_\textrm{I}$ quickly. 
We can therefore integrate out the heavy field $\chi$ as compared to the 
$\phi$ field having smaller mass and write down the 
effective potential during inflation in terms of $\phi$ only as given by
\begin{eqnarray}
V_{\rm{Inf}} & \simeq & \frac{1}{2} m^2 \phi^2 -\frac{c_1^2}{4 \lambda_{\chi}}\phi^4\label{veff}, \\
& = & M_P^4 \left [ \frac{1}{2} \tilde{m}^2 \tilde{\phi}^2 \left(1 - \alpha \tilde{\phi}^2 \right) \right]\label{veff1}.
\end{eqnarray}
For convenience the notations $\tilde{m}$ and $\tilde{\phi}$ are used to express $m$ and $\phi$ in terms of $M_P$ unit respectively. We will  describe our 
findings in $M_P$ unit in the rest of our discussion involving inflation. The parameter $\alpha$ is defined as 
$\alpha = \frac{c_1^2}{2 \lambda_{\chi} \tilde{m}^2}$. We assume $\alpha\tilde{\phi}^2 \ll 1$ so that this correction 
term does not deform the standard chaotic inflation model much.

The slow roll parameters are obtained from the standard definitions as given by 
\begin{equation}
\epsilon =\frac{1}{2}\Big(\frac{V_{\rm{Inf}}^{\prime}}{V_{\rm{Inf}}}\Big)^2 = 
\frac{2}{\tilde{\phi}^2}  
\Big[ \frac{1-2 \alpha \tilde{\phi}^2}{1-\alpha \tilde{\phi}^2} \Big]^2, \textrm{  }\eta =\frac{V_{\rm{Inf}}^{\prime\prime}}
{V_{\rm{Inf}}} = 
\frac{2}{\tilde{\phi}^2}  
\Big[ \frac{1-6 \alpha \tilde{\phi}^2}{1-\alpha \tilde{\phi}^2} \Big].
\end{equation}
The number of $e$-foldings is given by 
\begin{align}
N_e=\int_{\tilde{\phi}_{end}}^{\tilde{\phi}^*} 
\frac{\tilde{\phi} \left (1-\alpha \tilde{\phi}^2 \right)}{2(1-2 \alpha \tilde{\phi}^2)} 
d\tilde{\phi},
\end{align}
where $\tilde{\phi}^*$ and $\tilde{\phi}_{end}$ correspond to the field values 
at the point of horizon exit and end of inflaton respectively.  
The spectral index and tensor to scalar ratio are given by 
$n_s= 1-6\epsilon+2\eta$ and $r=16\epsilon$. The curvature perturbation is defined as
\begin{align}\label{PS}
P_s=\frac{V_\textrm{Inf}}{24\pi^2\epsilon}=\frac{\tilde{m}^2\tilde{\phi}^4}{96\pi^2}\frac{(1-\alpha\tilde{\phi}^2)^3}{(1-2\alpha\tilde{\phi}^2)^2}.
\end{align}
Observational value of $P_s$ is found to be $2.2\times 10^{-9}$ at the pivot scale $k^*\sim 0.05$ Mpc$^{-1}$\cite{Ade:2015xua}.
\begin{table}[H]
\begin{center}
\begin{tabular}{ |c | c | c | c |}
\hline
  $\tilde{m}$ & $\alpha$ & $n_s$ & $r$ \\
\hline
$5.83\times10^{-6}$ & $7\times10^{-4}$ & 0.966 & 0.097\\ 
$5.72\times10^{-6}$ & $9\times10^{-4}$ & 0.964 & 0.086\\
$5.59\times10^{-6}$ & $1.1\times10^{-3}$ & 0.962 & 0.076\\
$5.42\times10^{-6}$ & $1.3\times10^{-3}$ & 0.959 & 0.066\\
\hline
\end{tabular}
\caption{Inflationary predictions for $n_s$ and $r$ for different values of $\alpha$ with $N_e=60$.}
\label{tab:1}
\end{center}
\end{table}

We perform a scan over the parameters $m$ and $\alpha$ involved in Eq.(\ref{veff1}) so as to obtain $r$ and $n_s$ within the allowed range of 
\begin{figure}[h]
\begin{center}
\includegraphics[width=10cm, height=7cm]{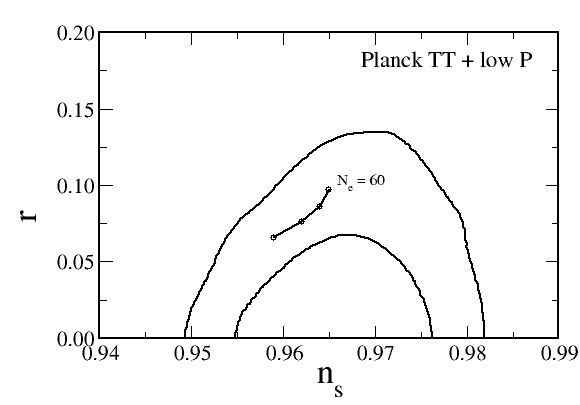}
\end{center}
\caption{Predictions for $n_s$ and $r$ (shown in Table \ref{tab:1}) as obtained from the deformed version of standard chaotic inflation
 are indicated by dark dots with $N_e=60$. A solid line joining them represents the predictions
 for $n_s$ and $r$ while $\alpha$ is varied. $1\sigma$ and $2\sigma$ contours of $n_s-r$ as obtained from Planck 2015 data\cite{Ade:2015xua} are included for reference.}
 \label{fig:planck}
\end{figure}
Planck 2015 \cite{Ade:2015xua}. We take $N_e$ as 60. Few of our findings for $n_s$ and $r$ in terms of the parameters $m$ and $\alpha$ are 
provided in the Table \ref{tab:1}. In Fig.\ref{fig:planck} we show our predictions for $n_s$ and $r$ by four dark dots joined by a line.
The dark dots represent the four sets of parameters mentioned in Table \ref{tab:1}. Along the line joining these, the parameter
 $\alpha$ is varied and correspondingly the magnitude of $\tilde{m}$ is adjusted mildly (as seen from Table \ref{tab:1}) in order to keep the curvature perturbation unchanged.
 The 1$\sigma$ and 2$\sigma$ contours from the Planck 2015\cite{Ade:2015xua} are also depicted as reference in Fig.\ref{fig:planck}. As an example with $\alpha\sim 7
\times 10^{-4}$ and $N_e=60$, we find $\tilde{\phi^*}\simeq 14.85$ (inflaton field value at horizon exit) and $\tilde{\phi}_{end}\simeq \sqrt
{2}$ (field value at the end of inflation). Hence the slow roll parameters $\epsilon$ and $\eta$ can be obtained at
 $\tilde{\phi}=\tilde{\phi}^*$ and we can determine $n_s$ and $r$. The parameter $m$ is fixed by the required
 value of curvature perturbation  $P_{s}=2.2\times 10^{-9}$\cite{Ade:2015xua}. $\tilde{m}$ is found to be $5.83\times 10^{-6}$ for the
 above values of $\alpha$ and $N_e$. As expected we obtain a smaller value of $r\sim 0.097$ compared to the standard chaotic 
inflation with $r\sim 0.133$ (and $n_s\simeq 0.966$)  as seen from Table \ref{tab:1} (first set).
\begin{figure}[H]
\begin{center}
\includegraphics[width=10cm, height=8cm]{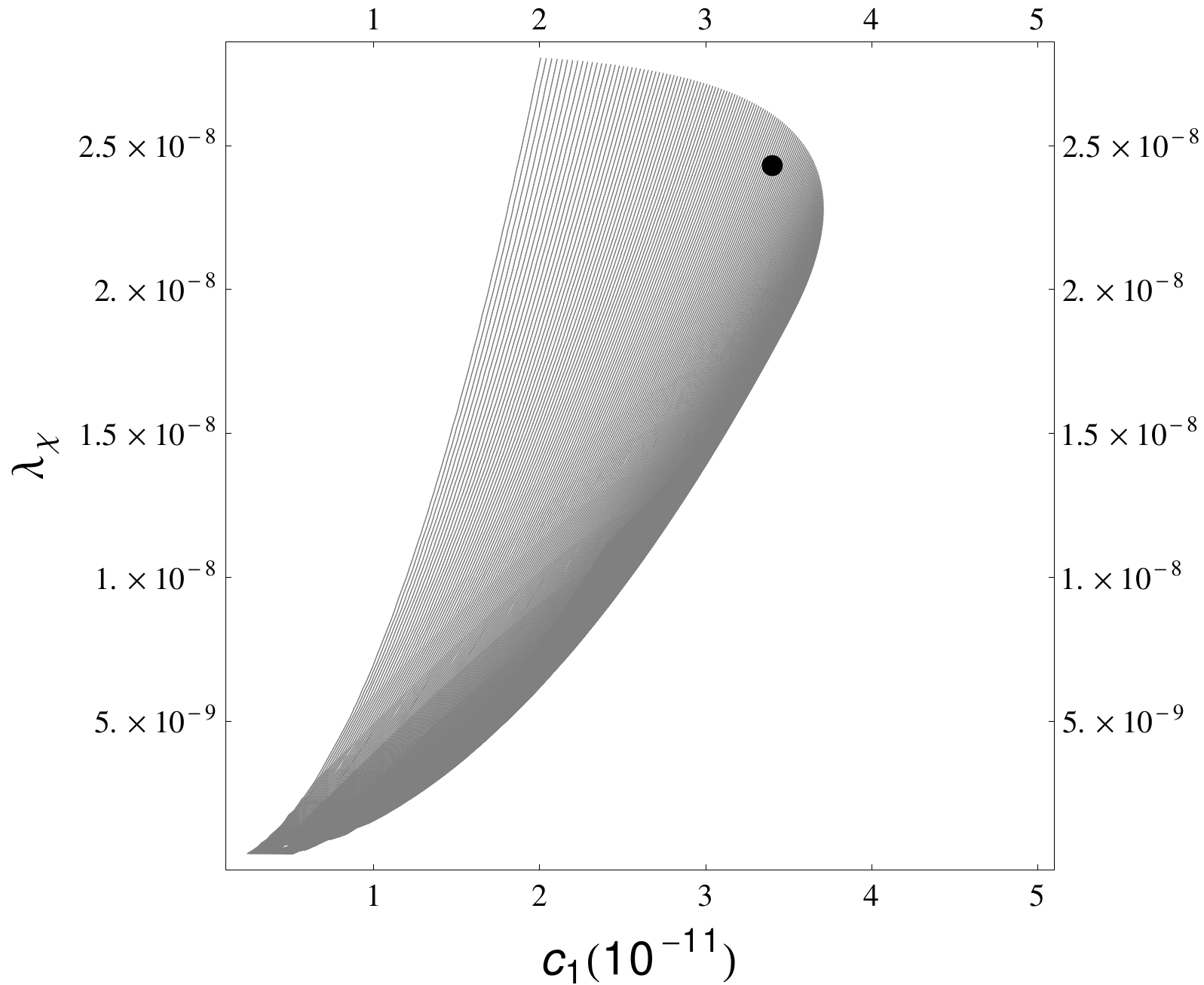}
\end{center}
\caption{Allowed region of $c_1$ (in unit of $10^{-11}$) and $\lambda_\chi$ is indicated by the shadowed region where the constraints [\textrm{(i)-(v)}] are included. The black dot represents the reference value of $c_1\simeq \tilde{m}^2 =3.39\times 10^{-11}$ and $\lambda_\chi\simeq 2.43\times10^{-8}$. $\alpha$ satisfies $2.7\times10^{-4}<\alpha<2.5\times10^{-3}$.}
\label{fig:cont1}
\end{figure}

Let us now proceed to determine the allowed range of parameters $c_1$ and $\lambda_\chi$ from estimate 
of $\alpha$ and $m$ we found above.
For this purpose, we first summarize the relevant points already  discussed. Specifically, 
during inflation:\textrm{(i)} $m_\chi^2 > H_{\textrm{Inf}}^2$ ,which indicates $c_1/\tilde{m}^2>1/12$ or equivalently
  $\lambda_\chi > \frac{\tilde{m}^2}{288\alpha}$. \textrm{(ii)} We assume $\chi$ field as sub-Planckian at the onset of inflation and 
afterwards. Hence $\tilde{\chi}_{\textrm{I}} \simeq (c_1/\lambda_\chi)\tilde\phi < 1$. \textrm{(iii)} As explained below in Eq.(6) 
we consider $\alpha\tilde\phi^2 < 1$.\textrm{(iv)} However note that this term should be sufficiently large so as to produce significant 
(at least ten percent) change in $r$ as compared to the minimal chaotic inflation. 
Ten percent or more reduction of $r$ can be achieved with $\alpha \tilde\phi^2\geq 0.05$. Therefore $\alpha$ is bounded 
by the inequality $0.05 \leq \alpha\tilde\phi^2 < 1$. 
Note that one can find $\tilde\phi^*(\alpha)$ (as function of $\alpha$) by solving $N_e = 60$ for a specific choice
of $\alpha$. Using the fact that we only keep terms of the order
 $\alpha \tilde\phi^2$ (i.e neglecting higher order terms), a suitable upper value of $\alpha \tilde\phi^2$ can
 be chosen as $ \alpha \tilde\phi^2 < 0.4$. Then plugging $\tilde{\phi}^*$ as a
function of $\alpha$, we obtain  $2.7\times 10^{-4}\leq \alpha \leq 2.5\times 10^{-3}$  as shown in Fig. \ref{fig:cont1}. To be concrete for the sake of discussion 
we choose a particular value of $\alpha$ within this range, say 
$\alpha = 7\times10^{-4} $ (see the first set of Table 1). One such $\alpha$ corresponds to a particular $\tilde m$ 
 value through 
Eq.(\ref{PS}) to have $ P_s =2.2\times 10^{-9} $ while $\tilde{\phi}^*$ is replaced by $\alpha-$dependence. Hence condition (i) can be translated 
as $\lambda_\chi > 1.68\times10^{-10}$ for $\alpha = 7\times10^{-4}$.
\textrm{(v)} An upper bound of $\lambda_\chi$ can be set from the requirement that involves the initial condition problem. Note
 that the universe during
the inflation is expected to be dominated by the $\phi$ field and hence $\lambda_\chi\chi^4 /4$ term should be 
sub-dominated 
compared to $(1/2)m^2\phi^2$ while initial $\tilde\phi^*$ can be large enough during the Planckian time and $\tilde\chi^*$ 
(initial value of $\chi$ before inflation starts) should remain 
sub-Planckian. For example considering $\tilde\phi^*\simeq 16$ and $\tilde\chi^*\simeq0.9$, we have 
$\lambda_\chi < 2.7\times10^{-8}$
such that $\lambda_\chi \tilde\chi^4/4 < (1/2) \tilde m ^2 \tilde\phi^2$. So $\lambda_\chi$ is restricted 
by $1.67\times 10^{-10} <\lambda_\chi< 2.7\times 10^{-8}$. From this we note that a choice $c_1 = \tilde m^2$ falls in the right ballpark which
 corresponds to $\lambda_\chi = (c_1/2\alpha) =2.43\times 10^{-8} $.
Considering the range of $\alpha$ as obtained, the allowed parameter space for $c_1$ and $\lambda_\chi$ are
 shown in Fig.\ref{fig:cont1} where the point corresponding to
$c_1 = \tilde m^2$ and $\lambda_\chi=2.43\times 10^{-8}$ is denoted by a dark dot.

We now turn our attention to the other part of the work which involves the SM Higgs doublet $H$ and its interaction with the
 inflation sector. The relevant tree level potential is given by
\begin{align}
V_{\textrm{II}}=\lambda_H\Big(H^\dagger H-\frac{v^2}{2}\Big)^2+\frac{\lambda_{\chi H}}{2}(\chi^2-v_\chi^2)\Big(H^\dagger H-\frac{v^2}{2}\Big)+\frac{\lambda_{\phi H}}{2}\phi^2\Big(H^\dagger H-\frac{v^2}{2}\Big).
\end{align}
As discussed earlier we expect the threshold effect on the running of the Higgs quartic coupling to appear from its
 interaction with the $\chi$ field only as $\langle\chi\rangle$ can be large while $\langle\phi\rangle=0$. Hence we drop the last term involving interaction between $H$ and 
inflaton $\phi$ for the rest of our discussion by assuming $\lambda_{\phi H}$ vanishingly small. To explore the 
stability of the Higgs potential we need to consider the $\lambda_\chi$ term from Eq.(\ref{infpot}) as well. The part of the entire 
potential $V_\textrm{I}+V_{\textrm{II}}$ relevant to discuss the vacuum stability issue is therefore given by
\begin{align}\label{totV}
V_0=\lambda_H\Big(H^\dagger H-\frac{v^2}{2}\Big)^2+\frac{\lambda_{\chi H}}{2}(\chi^2-v_\chi^2)\Big(H^\dagger H-\frac{v^2}{2}\Big)+\frac{\lambda_\chi}{4}(\chi^2-v_\chi^2)^2.
\end{align}
\noindent The minimum of $V_0$ is given by
\begin{align}\label{min1}
\langle H^\dagger H\rangle=\frac{v^2}{2} \textrm{,    }\langle\chi\rangle=v_\chi,
\end{align}
where we have considered $\lambda_H$, $\lambda_\chi>0$. Note that the above minimum of $V_0$ (the EW minimum) corresponds 
to vanishing vacuum energy {\it{i.e.}} $V_0^{\textrm{EW}}=0$. Now in order to maintain the stability of the potential, $V_0$ should remain 
positive ($V_0>0$) even when the fields involved ($\chi$ and $H$) are away from their respective values at the EW minimum.
 Since the couplings depend on the renormalization scale $\mu$ $(\sim$ field value) we must 
ensure $\lambda_\chi(\mu)$, $\lambda_H(\mu)>0$ in order to avoid any deeper minimum (lower than the $V_0^{\textrm{EW}}$) away from 
the EW one.

In order to study the running of all the couplings under consideration, we consider the renormalization group (RG) equations for them.
 Below we provide the RG equations for $\lambda_H$, $\lambda_{\chi H}$ and $\lambda_\chi$\cite{EliasMiro:2012ay} as
\begin{align}
\frac{d \lambda_H}{dt}=&\beta^{SM}_{\lambda_H}+\frac{1}{16 \pi^2}\lambda_{\chi H}^2\label{RG},\\
\frac{d \lambda_{\chi H}}{dt}=&\frac{1}{16 \pi^2}\Big\{12 \lambda_H\lambda_{\chi H}+8\lambda_{\chi}\lambda_{\chi H}+4\lambda_{\chi H}^2+6 y_t^2\lambda_{\chi H}-\frac{3}{2}g_1^2\lambda_{\chi H}-\frac{9}{2}g_2^2\lambda_{\chi H}\Big\}\label{RG1},\\
\frac{d \lambda_{\chi}}{dt}=&\frac{1}{16 \pi^2}\Big\{20 \lambda_{\chi}^2+2\lambda_{\chi H}^2\Big\}\label{RG2}.
\end{align}
$\beta^{SM}_{\lambda_H}$is the three loop $\beta$-function for the Higgs quartic coupling\cite{EliasMiro:2012ay} in SM, which is
 corrected by the one-loop contribution in presence of the $\chi$ field. The RG equation for two new physics parameters 
$\lambda_{\chi H}$ and $\lambda_\chi$ are kept at one loop. The presence of the $\chi$ field is therefore expected to modify the stability
 conditions above its mass $m_\chi$.

Apart from the modified  running of the Higgs quartic coupling, vacuum stability is also affected by the presence of the threshold correction from the 
heavy $\chi$ field which carries a large vev. The mass of the $\chi$ field is given by $m_\chi=\sqrt{2\lambda_\chi}v_\chi$ 
(see Eq.(\ref{chiv}) with $\phi=0$ after inflation), the heavy field $\chi$ can be integrated out below $m_\chi$. By solving the equation of motion of $\chi$ 
field we have
\begin{align}\label{xvev}
\chi^2\simeq v_\chi^2-\frac{\lambda_{\chi H}}{\lambda_\chi}\Big(H^\dagger H-\frac{v^2}{2}\Big).
\end{align}
Hence below the scale $m_\chi$, the effective potential of $V_0$ becomes
\begin{align}\label{eff}
V_0^{\textrm{eff}}\simeq \lambda_h \Big(H^\dagger H-\frac{v^2}{2}\Big)^2, \textrm{ with } \lambda_h({m_\chi})=\Big[\lambda_H-\frac{\lambda_{\chi H}^2}{4\lambda_\chi}\Big]_{m_\chi},
\end{align}
where Eq.(\ref{xvev}) is used to replace $\chi$ into Eq.(\ref{totV}). Therefore below $m_\chi$, $\lambda_h$ corresponds to the SM Higgs quartic 
coupling and above $m_\chi$ it gets a positive shift $\delta \lambda=\frac{\lambda_{\chi H}^2}{4\lambda_\chi}$. This could 
obviously help in delaying the Higgs quartic coupling to become negative provided $m_\chi$ is below the SM instability scale, {\it{i.e.}} $m_\chi<\Lambda_I^{\textrm{SM}}$. In this analysis, we investigate
 the parameter space for which the Higgs quartic coupling remains positive upto the scale $M_P$. This however depends 
upon $m_\chi$ and $\delta \lambda$. Note that $\lambda_\chi$ involved in both $m_\chi$ and $\delta \lambda$ which is somewhat 
restricted from inflation (see Fig.\ref{fig:cont1}). On the other hand $v_\chi$ is not restricted from inflation. We will have
 an estimate of $v_\chi$ by requiring $m_\chi<\Lambda_I^{\textrm{SM}}$ (using a specific $\lambda_\chi$ value corresponding to set-1 of Table 1.) We also consider
 $\delta \lambda\sim \lambda_h({m_\chi})$ to avoid un-naturalnes in the amount of shift. This consideration in turn fixes $\lambda_{\chi H}$.

Once the electroweak symmetry is broken, the diagonalization of the mass-square matrix involving quadratic terms of $\chi$, $H$ and their mixing ($\lambda_{\chi H}$ term) yields one light and one 
heavy scalars. Using the unitary gauge for redefining the Higgs doublet as $H^T=\Big(0,\frac{v+h}{\sqrt{2}}\Big)$, the masses 
associated with the light and heavy eigenstates are  
\begin{align}
m^2_{\textrm{$h$,$h_H$}}=\Big[\lambda_H v^2+\lambda_{\chi}v_{\chi}^2\mp\sqrt{\Big(\lambda_Hv^2-\lambda_{\chi}v_{\chi}^2\Big)^2+\lambda_{\chi H}^2v^2 v_{\chi}^2} \Big],
\end{align}
where $h$ and $h_H$ correspond to light and heavy Higgs. In the limit of small mixing angle $\theta=\Big[\frac{1}{2}\textrm{tan}^{-1}\frac
{\lambda_{\chi H}v_{\chi}v}{\lambda_H v^2-\lambda_{\chi}v_{\chi}^2}$\Big], $m_{h}$ 
becomes $m_{h}\simeq \sqrt{2} v\Big(\lambda_H-\frac{\lambda_{\chi H}^2}{4 \lambda_\chi}\Big)$ with $\lambda_\chi v_\chi^2\gg \lambda_Hv^2$. In order to avoid the unwanted negative
 value of $m_{h}$, the extra stability condition $4\lambda_\chi \lambda_H>\lambda_{\chi H}^2$ should be maintained (with $\lambda_{\chi H}>0$).
 However it is pointed out in \cite{EliasMiro:2012ay} that it is sufficient that this condition should be satisfied for a short interval 
around $m_\chi$ for $\lambda_{\chi H}>0$. However for $\lambda_{\chi H}<0$, this extra stability condition becomes
 $2\sqrt{\lambda_\chi(\mu)\lambda_H(\mu)}+\lambda_{\chi H}(\mu)>0$. As found in \cite{EliasMiro:2012ay}, it is difficult
 to achieve the absolute stability of $V_0$ till $M_P$ in this case. We restrict ourselves into the case $\lambda_{\chi H}>0$ for the present work. 

\begin{table}[H]
\begin{center}
\begin{tabular}{|c | c | c | c | c | c |}
\hline
Scale ($\mu$)& $y_t$ & $g_1$ & $g_2$ & $g_3$ & $\lambda_h$\\
\hline
$m_t$ & $0.93668$ & $0.357632$ & $0.648228$ & $1.166508$ & $0.127102$\\
\hline
\end{tabular}
\end{center}
\caption{Values of couplings in SM at $m_t=173.3$ GeV.}
\label{tab:ini}
\end{table}

\begin{figure}[H]
\begin{center}
\includegraphics[width=10cm, height=7cm]{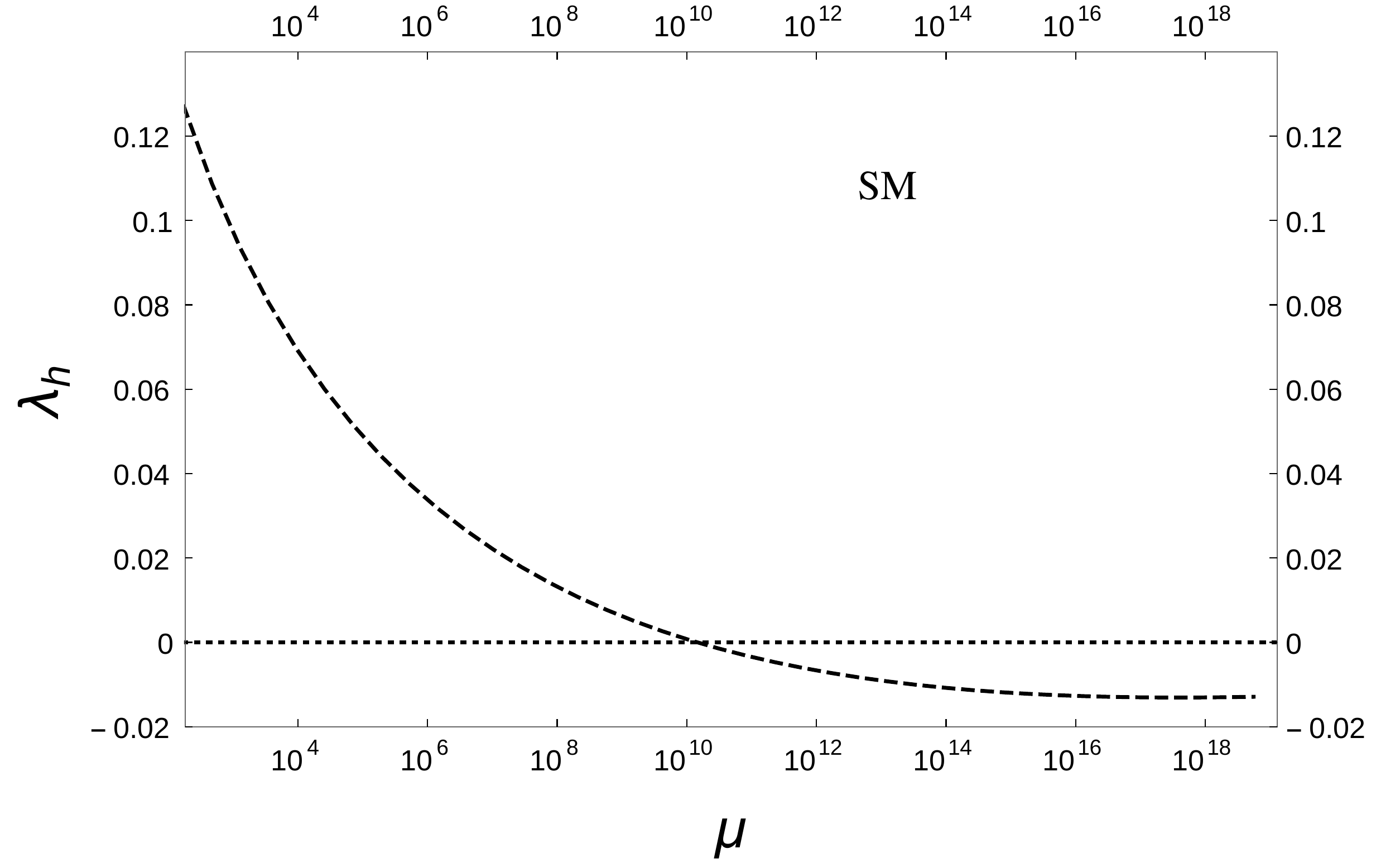}
\end{center}
\caption{Running of the SM Higgs quartic coupling till $M_P$ is shown for $m_t=173.3$ GeV, $m_h=125.66$ GeV and $\alpha_s(M_Z)=0.1184$.}
\label{fig:SM}
\end{figure}

To have a concrete understanding of the vacuum stability issue in this set-up, we first estimate several parameters at a 
scale of top quark mass $m_t=$173.3 GeV. The values of the top quark Yukawa coupling ($y_t$), gauge couplings ($g_i$) and Higgs quartic
 coupling $\lambda_h$ are taken at two loop NNLO precision following \cite{Buttazzo:2013uya}. These are mentioned in Table \ref{tab:ini}. 
We then use the RG equations (Eqns.(13-15)) for these parameters to study the running of $\lambda_h$ as shown in Fig. \ref{fig:SM}. We also consider the three loop SM RGE for the gauge couplings.
The instability scale then turns out to be $\Lambda_I^{\textrm{SM}}\simeq 1.2\times 10^{10}$ GeV for $m_t=173.3$ GeV, $m_h=125.66$ GeV and $\alpha_s(M_Z)=0.1184$.

\begin{table}[H]
\begin{center}
\begin{tabular}{|c | c | c | c | c | c | c | c |}
\hline
Scale($\mu$) & $\lambda_{\chi}$ & $\lambda_{\chi H}$ & $y_t$ & $g_1$ & $g_2$ & $g_3$ & $\lambda_H$\\
\hline
$m_\chi$ & $2.38\times10^{-8}$ & $3.66\times 10^{-5}$ & $0.59523$ & $0.386864$ & $0.58822$ & $0.72327$ & $0.0278$\\
\hline
$M_P$ & $2.45\times 10^{-8}$ & $3.56\times 10^{-5}$ & $0.39112$ & $0.467056$ & $0.509155$ & $0.49591$ & $\mathcal{O}(10^{-5})$\\
\hline  
\end{tabular}
\end{center}
\caption{Values of couplings at $m_{\chi}$ and $M_P$ for $m_{\chi}=8\times 10^{7}$ GeV and $m_t=173.3$ GeV. }
\label{tab:Pos}
\end{table}

Let us now proceed to the case with SM+Inflation extension. In this case, above the energy scale $m_\chi$, two other couplings $\lambda_\chi$ 
and $\lambda_{\chi H}$ will appear as in Eq.(\ref{totV}). 
As discussed earlier, we already have an estimate of $\lambda_\chi$ to 
have successful results in inflation sector with $c_1=\tilde{m}^2$. 
\begin{figure}[h]
\begin{center}
\includegraphics[width=10cm, height=7cm]{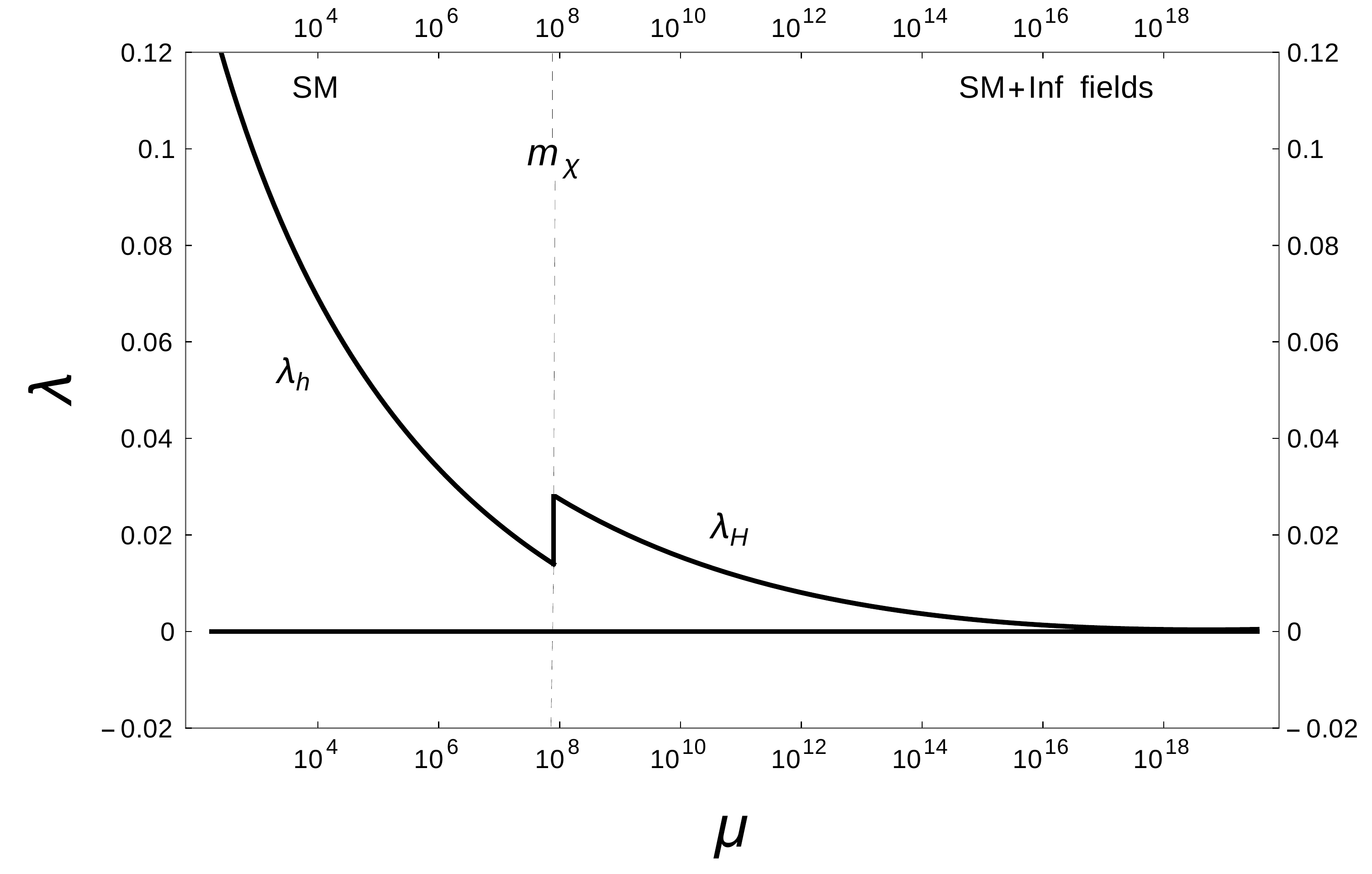}
\end{center}
\caption{Running of Higgs quartic coupling till $M_P$ for $\lambda_{\chi H}({m_\chi})=3.66\times 10^{-5}, \lambda_{\chi}({m_\chi})=2.38\times 10^{-8}$ with $m_{\chi}=8\times 10^{7}$ GeV.} 
\label{fig:ze}
\end{figure}
We consider the corresponding value of $\lambda_\chi=2.43\times 10^{-8}$ at inflation scale 
$\Lambda_{\textrm{Inf}}\sim V_{\textrm{Inf}}^{1/4}\simeq10^{16}$ GeV. The initial value of $\lambda_\chi$ should be fixed at $m_\chi$ 
(remains same at $m_t$) in such a way that it can reproduce $\lambda_\chi$($\Lambda_\textrm{Inf}$) correctly through its RG equation in Eq.(\ref{RG2}). 
$\lambda_{\chi H}$ $(m_\chi)$ is chosen to achieve a natural enhancement $\delta \lambda\sim\lambda_h(m_\chi)$ at $m_\chi$. Hereafter above 
$m_\chi$, the Higgs quartic coupling $\lambda_H$ is governed through the modified RG equation\footnote{It can be noted that such a $\lambda_{\chi H}$ does not alter the running of $\lambda_\chi$ much.} as in Eq.(\ref{RG}). Note that even if $\lambda_\chi$
 is known it does not fix $m_\chi$($=\sqrt{2\lambda_\chi}v_\chi$) completely. Therefore we can vary $v_\chi$
to have $m_\chi<\Lambda_I^{\textrm{SM}}$.

 \begin{figure}[h]
\centering
 \begin{subfigure}{.5\textwidth}
 \centering
 \includegraphics[width=7.5cm, height=6cm]{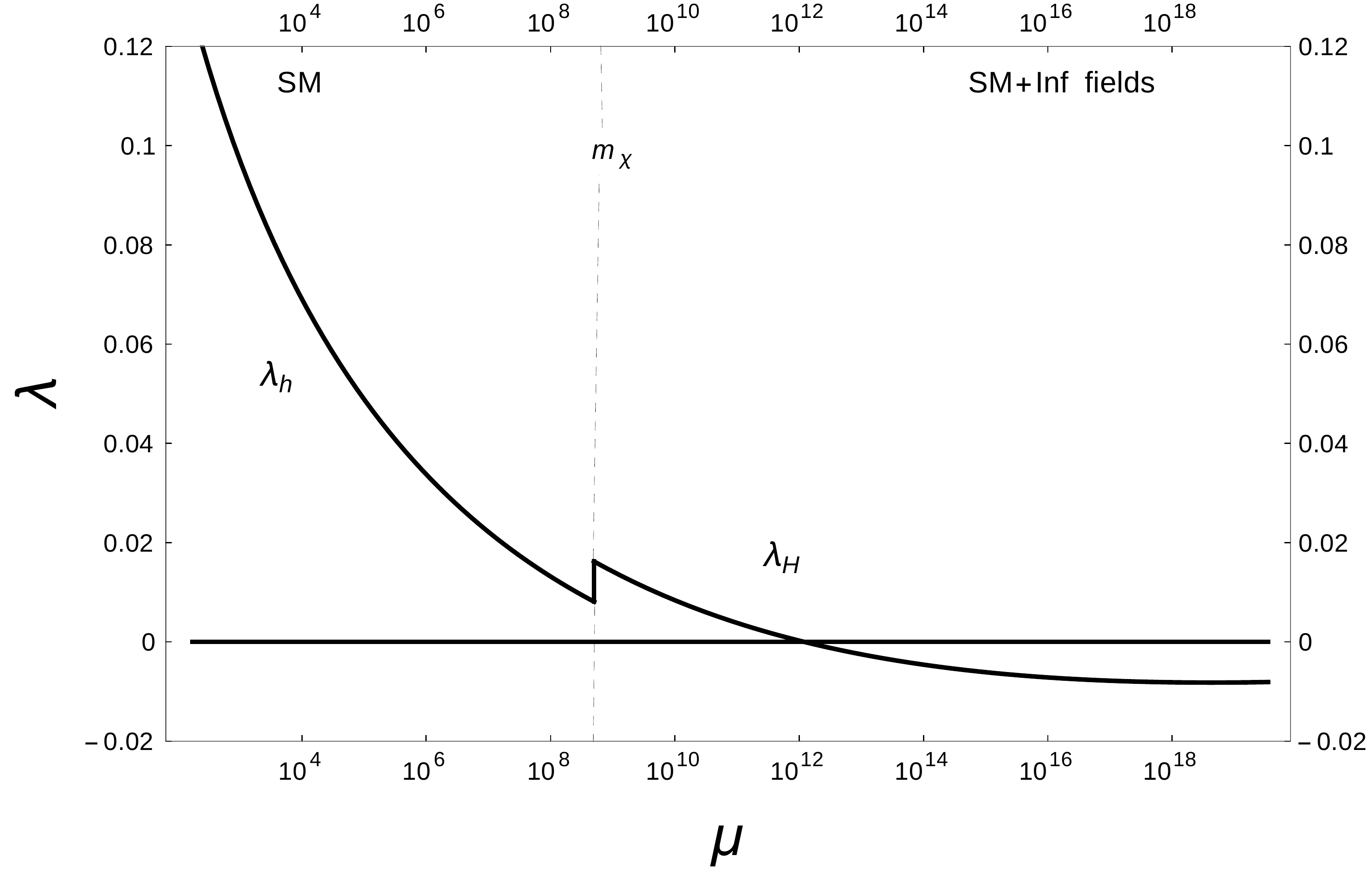}
 \caption{ }
 \label{sub1}
 \end{subfigure}
 \begin{subfigure}{.4\textwidth}
 \centering
\includegraphics[width=7.5cm, height=6cm]{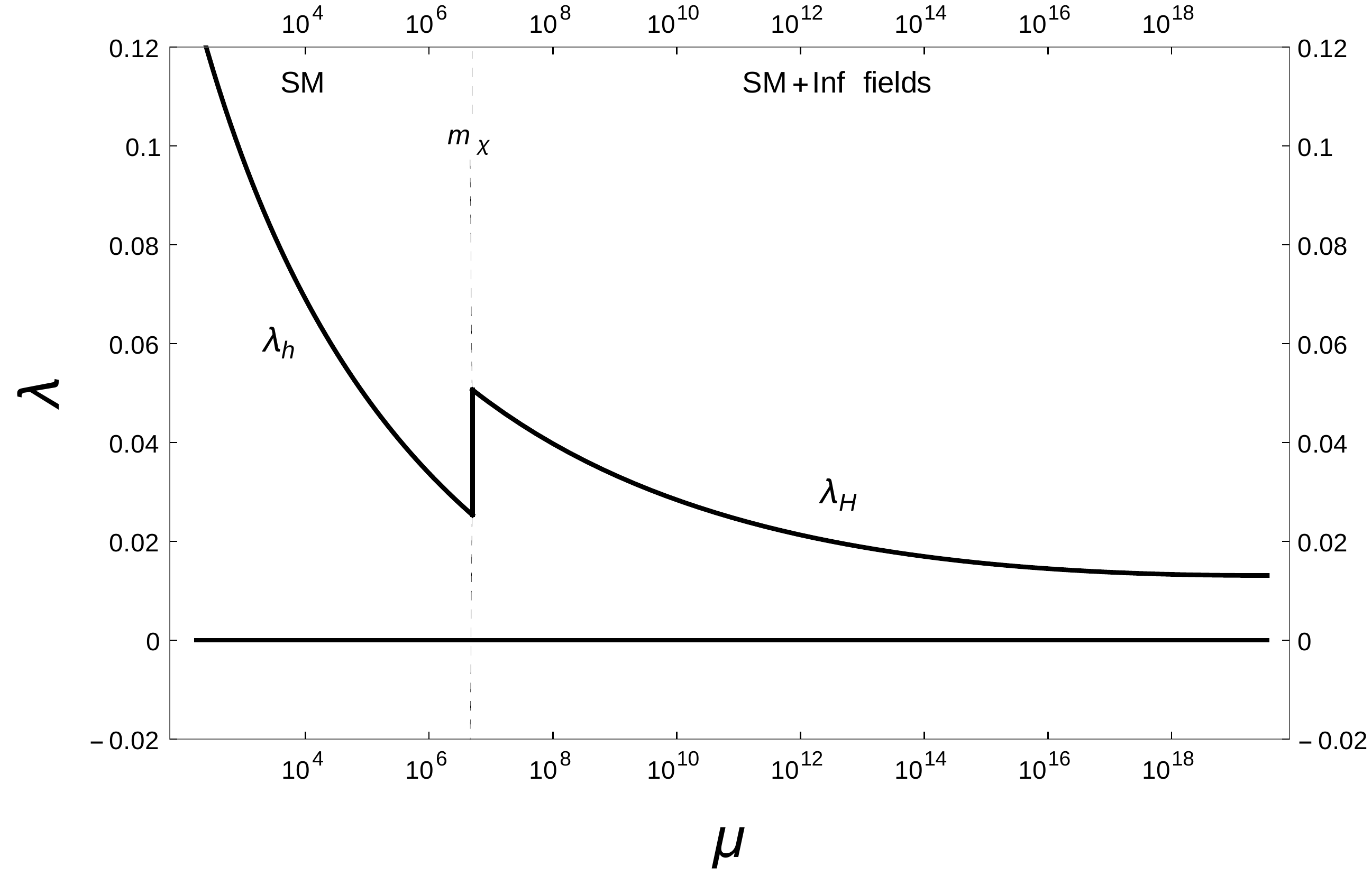}
 \caption{ }
 \label{sub2}
 \end{subfigure}
 \caption{Running of Higgs quartic coupling till $M_P$ for (a) $m_{\chi}(=5\times10^8 \textrm{ GeV})>m_C$, (b) $m_{\chi}(=5\times10^6  \textrm{ GeV})<m_C$. We consider $\delta\lambda=\lambda_h(m_\chi)$ which corresponds to the choice for (a) $\lambda_{\chi H}({m_\chi})=2.8\times 10^{-5}$, $\lambda_{\chi}({m_\chi})=2.4\times10^{-8}$ and (b) $\lambda_{\chi H}({m_\chi})=4.88\times 10^{-5}, \lambda_{\chi}({m_\chi})=2.35\times10^{-8}$ respectively.}
\label{fig:po}
\end{figure}
\begin{figure}[H]
\begin{center}
\includegraphics[width=10cm, height=6.5cm]{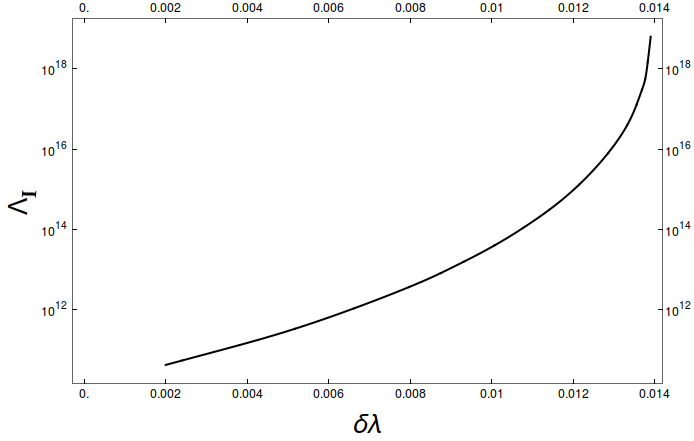}
\end{center}
\caption{Variation of instability scale $\Lambda_I$ with change of $\delta\lambda$ where $m_\chi=m_C\simeq 8\times10^7$ GeV.} 
\label{fig:changeI}
\end{figure}
\begin{figure}[H]
\begin{center}
\includegraphics[width=10cm, height=6.5cm]{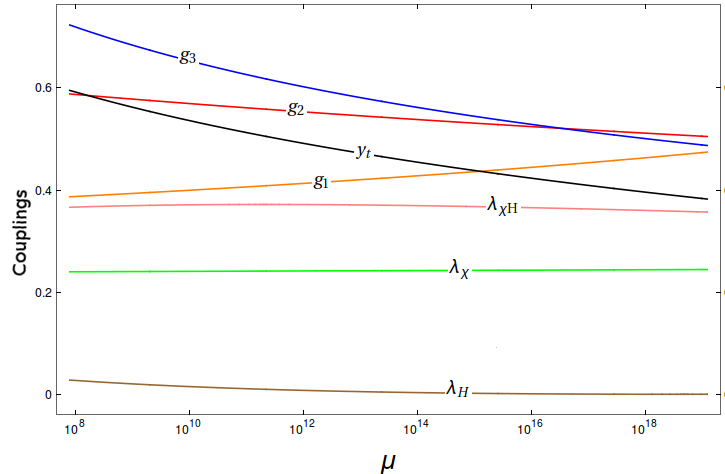}
\end{center}
\caption{Running of the couplings $g_{i=1,2,3}$, $y_t$, $\lambda_H$, $\lambda_{\chi H}$ (in unit of $10^{-4}$) and $\lambda_\chi$ (in unit of $10^{-7}$) in SM+Inflation scenario from $m_\chi=m_C\sim 8\times 10^7$ GeV to $M_P$.} 
\label{fig:all}
\end{figure}

With the above mentioned scheme, we study the running of the Higgs quartic coupling for different $m_\chi$ satisfying 
$m_\chi<\Lambda_I^{\textrm{SM}}$. We find that with $m_\chi=8\times 10^{7}$ GeV, $\lambda_H$ becomes vanishingly small at $M_P$ (hence the new instability scale $\Lambda_I$ becomes $\sim M_P$) as shown in 
Fig.\ref{fig:ze}. We specify the corresponding $m_\chi$ value as $m_C(=8\times10^7$ GeV). The other relevant couplings at $m_\chi$ and
 at $M_P$ are given in Table \ref{tab:Pos}. 
It is then observed that in order to keep the Higgs quartic coupling positive all the way upto $M_P$, we should ensure $m_\chi<m_C$. For example with
 $m_\chi=5\times 10^8$ GeV ($>m_c$) we see $\lambda_H$ becomes negative at scale $\sim 10^{12}$ GeV as shown in 
 Fig.\ref{sub1}. On the other hand in Fig.\ref{sub2} it is seen that $\lambda_H$ remains positive till $M_P$ for $m_\chi\sim 5\times 10^6$ GeV $<m_C$. In doing so we consider the amount of positive shift at $m_\chi$ to be defined with
$\delta\lambda\sim \lambda_h(m_\chi)$. In Fig.\ref{fig:changeI} we provide the variation of instability scale $\Lambda_I$ (which is atmost $\sim M_P$ for $m_\chi=m_C$ case) in SM+Inflation extension if we relax this assumption by changing
 $\delta \lambda$ arbitarily. To give a feeling about how other couplings
  are changing with $\mu$, we plot running of gauge couplings $g_{i=1,2,3}$, top quark yukawa coupling $y_t$, $\lambda_\chi$, $\lambda_H$ 
  and $\lambda_{\chi H}$ in Fig.\ref{fig:all} with $m_\chi=m_C=8\times 10^7$ GeV.

For completeness we now comment on reheating in the present set-up. Once the inflation ends, the inflaton $\phi$ will oscillate around the minimum at
$\phi=0$. The decay of $\phi$ would then proceed provided it interacts with the SM fields. Here we do not attempt to discuss it in details. Instead we only mention about the possibilities. The details will be discussed elsewhere in a future study. 
We may consider terms in the Lagrangian as 
\begin{align}
-\mathcal{L}_{int}=y_\phi\phi NN+y_\nu \bar{L}\tilde{H} N,
\end{align}
where $N$ is the right handed (RH)  neutrino and $L$ is the SM lepton doublet. $y_\phi$, $y_\nu$ are the respective couplings. Note that the first term is 
an explicit $Z_2$ breaking term and hence $y_\phi$ is expected to be small in the present set-up\footnote{ $y_\phi$ and $y_\nu$ should be sufficiently small so that it does not contribute to RG evolutions of all other couplings.}. The corresponding reheat temperature is then found 
 to be $T_r\simeq y_\phi\sqrt{\frac{m M_P}{8\pi}}\sim 10^{15} y_\phi$ GeV where we have used $m=1.4\times 10^{13}$ GeV from set-1, Table 1. A further decay of RH neutrinos into $L$ and $H$ can be responsible for leptogenesis
 \cite{Fukugita:1986hr,Flanz:1994yx, Covi:1996wh, Plumacher:1996kc, Buchmuller:1997yu}. The term $y_\phi \phi NN$ can not however provide
 the mass of the RH neutrinos as $\langle\phi\rangle=0$. A mass term like $M_NNN$ has to be present. If we turn our attention to the other field $\chi$ involved in
 the inflation sector, we note that this field will oscillate about $\langle\chi\rangle$ at the end of inflation. The decays of it can 
proceed via $\chi\rightarrow hh$ with $\Gamma_\chi=\frac{\lambda_{\chi H}^2 m_\chi}{256 \pi \lambda_\chi}$. Then two cases may arise;
(i) $\Gamma_{\phi}\ll\Gamma_{\chi}$: in this case the $\chi$ field will decay very fast. However reheat of universe will finish 
much later after $\chi$ field decay. So any radiation energy density produced by $\chi$ will be strongly diluted during the matter
 dominated phase governed by the oscillations of $\phi$. 
 (ii) $\Gamma_{\phi}\gg\Gamma_{\chi}$: note that energy density of the $\chi$ field is much less than that of $\phi$ and
  universe will reheat once decay of $\phi$ field is completed. Hence the completion of the inflaton decay into radiation, when Hubble becomes of
order  $\Gamma_{\chi}$, $\chi$ field will decay to radiation. So the remnant radiation in this case will be a mixture of $\phi$ and $\chi$
 products.

The discovery of the Higgs boson and the precise determination of its mass at LHC provide us an estimate of the 
Higgs quartic coupling in the SM. However the high scale behavior of this coupling, whether or not it becomes negative, poses
 plethora of questions in terms of the stabilization of the electroweak minimum. Though the present data favors the metastability of
 this vacuum, it is very much dependent on the precision of the top mass measurement. Furthermore inflation in the early universe
 provides additional threat as it can shift the Higgs field during inflation into the unwanted part of the Higgs potential and 
hence metastability can also be questioned. As a resolution to this, we propose introduction of the inflation sector 
consisting of two scalar fields $\phi$ and $\chi$ and 
their interaction with the SM Higgs. While $\phi$ is playing the role of the inflaton having the potential $m^2\phi^2/2$, the other 
field $\chi$ provides a deviation in terms of prediction of the spectral index and tensor to scalar ratio so has to be consistent 
with the Planck 2015 data. We have shown that the $\chi$-Higgs coupling can have profound effect in the running of Higgs quartic
 coupling considering the positive shift through the threshold effect at a scale $ m_\chi$. It turns out the quartic coupling
 of the $\chi$ field is restricted to achieve successful inflation. This in turn constrain the other new physics parameters space if we want
 to make the Higgs potential completely stable upto $M_P$. The scenario also alleviates the problem of instability of the EW vacuum 
during inflation as the Higgs field is stabilized at origin having a mass larger($\sim 100 H_{\textrm{Inf}}^2$) than the Hubble during
 inflation. Once the inflation is over, it smoothly enters into the near EW minimum. 
We have also commented on the possible reheating scenario in brief. The vev of the $\chi$ field breaks the $Z_2$ symmetry which may
spontaneously lead to domain wall problem. An explicit $Z_2$- symmetry breaking term or gauging the symmetry would help in 
resolving the issue. As an extension of our present set-up, one can possibly consider a $U(1)_{B-L}$ embedding of the entire framework where the 
neutrino masses and several other related aspects like effect of gauge bosons on running etc. can be simultaneously addressed.

\end{document}